\documentclass[aps,reprint,amsmath,amssymb]{revtex4-2}
\usepackage{graphics}
\usepackage{subfigure}
\usepackage{epsfig}
\usepackage{epsf,epic}
\usepackage{dcolumn}
\usepackage{color}
\usepackage{hyperref}
\usepackage{marvosym}
\usepackage{subfigure}
\usepackage{amsmath}
\usepackage{amssymb}
\usepackage{amsfonts}
\usepackage{wrapfig}
\usepackage{pstricks}
\usepackage{multirow}
\usepackage{bm}
\graphicspath{{./figs/}}
\usepackage{dcolumn}
\newcommand{\etal}{\textit{et al.\ }}
\newcommand{\ie}{\textit{i.e.\ }}

\begin{document}
\title{Spinel LiGa$_5$O$_8$ prospects as ultra-wideband-gap semiconductor: band structure, optical properties and doping.}

\author{Walter R. L. Lambrecht}\email{walter.lambrecht@case.edu}

\affiliation{Department of Physics, Case Western Reserve University, 10900 Euclid Avenue, Cleveland, Ohio 44106-7079, USA}

\begin{abstract}
  LiGa$_5$O$_8$ in the spinel type structure is investigated as a potential ultra-wide-band-gap  semiconductor. The band structure is determined using the
  quasiparticle self-consistent $GW$ method and the optical properties are calculated at the Bethe Salpeter Equation level including electron-hole interaction effects. The optical gap including exciton effects and an estimate of the zero-point motion electron phonon coupling renormalizations is estimated to be about 5.2$\pm0.1$ eV with an exciton binding energy of about 0.4 eV. Si doping as potential $n$-type dopant is investigated and found to be a promising  shallow donor.
\end{abstract}
\maketitle
\section{Introduction}
There is a need for ultra-wide-band-gap semiconductors  both for optoelectronic and
high-power applications. The high-power applications benefit from a large gap because it is found that the breakdown fields scale with the band gap.
Recently, there has been a lot of interest in $\beta$-Ga$_2$O$_3$, because of its
gap of $\sim$4.9 eV.\cite{Matsumoto74,Sasaki13} In spite of such a large gap it can be doped with Si, Ge or Sn to
give semiconducting properties. This finding has sparked a world-wide development of this material in epitaxial form.
The development of heterojunction devices involving even higher gap (Al$_x$Ga$_{1-x}$)$_2$O$_3$ is  making rapid progress.  For the higher Al-concentrations $n$-type doping is
difficult but $p$-type doping appears to be elusive so far  even for Ga$_2$O$_3$. This is attributed to formation of 
self-trapped hole polarons, related in turn to high valence band effective masses.
While there are some reports of low-level $p$-type doping by means of defect complexes,
these remain controversial and the hole mobility is very low. \cite{Chi23,Chikoidze19}
$\beta$-Ga$_2$O$_3$ also has led to a renewed interest in fundamental properties because of the
unusual monoclinic structure with both tetrahedral and octahedral Ga sites.

On the other hand, LiGaO$_2$ (lithium gallate) has an even wider band gap and was recently
also predicted to be $n$-type dopable by Si or Ge. \cite{Boonchun19,Dabsamut20}
Like Ga$_2$O$_3$ bulk crystals can be grown and offer the possibility of
homo-epitaxial growth.  It has a much simpler wurtzite based crystal structure with all Ga and Li in tetrahedral coordinations and can be thought of as
derived from ZnO by replacing the group-II ion Zn by alternating group-I (Li)
and group III Ga ions.  Further band gap tuning is possible by replacing Ga
by Al and Li by Na in the same $Pna2_1$ crystal structure or closely related structures.\cite{Popp22,Radha21ligao2} The anisotropy of the material  leads to interesting
exciton splittings with large exciton binding energies and recent calculations
\cite{Dadkhah23} were found to be in excellent agreement with experiment\cite{Trinkler17,Trinkler22,Tumenas17} and give an exciton gap of 6 eV.
It thus appears useful to develop epitaxial growth and doping of this material to pursue it as an active semiconductor material and not merely an optical material. 

\begin{figure}
  \includegraphics[width=6cm]{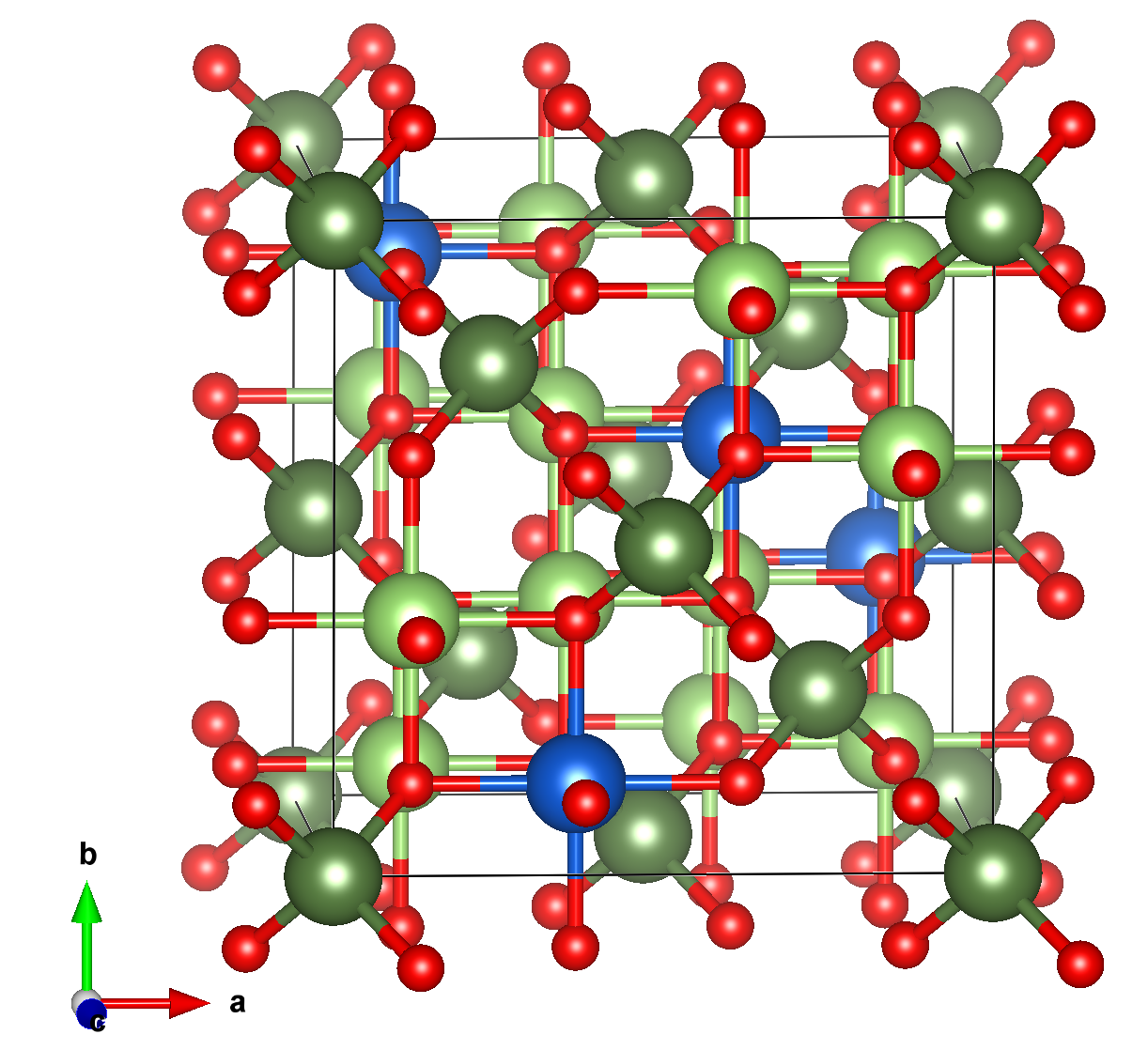}
  \caption{Crystal structure of LiGa$_5$O$_8$, blue spheres Li, light green, octahedral Ga, dark green tetrahedral Ga, red, O} \label{figstruc}
\end{figure}

In the Li-Ga-O ternary system, another stable compound exists with formula 
LiGa$_5$O$_8$. One can think of it as taking 4 LiGaO$_2$ units and replacing
3 Li with one Ga to maintain the charge balance. However, LiGa$_5$O$_8$ is known to have a different crystal structure. It has a cubic spinel-type structure with Ga occurring in both tetrahedral and octahedral sites and Li in octahedral sites.\cite{Joubert63} Its structure and related information can be found in \cite{liga5o8struc,Matproj}. We used the experimental crystal structure from the International Crystallographic Database (ICSD),\cite{icsdliga5o8} with cubic lattice constant of 8.203 \AA\  and space group (No. 212) $P4_332$ or $O^6$ 
as shown in Fig. \ref{figstruc}. Interestingly, a defective spinel phase also occurs in Ga$_2$O$_3$
as a metastable phase and has been labeled the $\gamma$-phase \cite{Mahitosh22}.
It has recently been observed to occur
during certain processing steps in doped $\beta$-Ga$_2$O$_3$ and a model for how the transition from
$\beta$ to $\gamma$ phase can proceed via a succession of defect formations was proposed by Hsien-Lien \etal
\cite{Hsien-Lien23,Hwang23}
Since LiGa$_5$O$_8$  is a compound with composition in between Ga$_2$O$_3$ and LiGaO$_2$ it is of interest to determine its band structure and optical properties to assess its potential as UWBG semiconductor. It has previously been studied as a phosphorescent material  when doped with Cr.\cite{DeClercq17,Sousa20,Huang2018}
Very recently, LiGa$_5$O$_8$ was grown epitaxially and reported to exhibit
$p$-type conduction.\cite{Zhao23}
The relation to the $\gamma$-phase of Ga$_2$O$_3$, which is known to have a higher band gap than the $\beta$-phase, provides additional interest. It suggests that adding a small amount of Li may stabilize the $\gamma$-phase. 

This paper presents results on the electronic band structure using
the state-of-the-art quasiparticle self-consistent $GW$ method and also
calculates optical properties, including excitonic effects. Since the unit cell already has 56 atoms, (8 formula units) we can already obtain an idea of doping by simply replacing one Ga by a Si. We thus also study
a Li$_4$Ga$_{19}$SiO$_{32}$ cell with Si placed on either a tetrahedral or octahedral Ga site.

\section{Computational Methods}
The calculations were performed using the {\sc Questaal} suite,\cite{questaalpaper}  which uses the full-potential linearized muffin-tin orbital
method(FP-LMTO) \cite{Kotani10} to implement both density functional theory (DFT) and many-body perturbation theory (MBPT)
approaches such as Hedin's $GW$ method for quasiparticle band structures  ($G$ for Green's function and $W$ for screened Coulomb interaction)\cite{Hedin65,Hedin69} in a quasiparticle self-consistent version called QS$GW$, \cite{MvSQSGWprl,Kotani07}
and the Bethe Salpeter Equation (BSE) approach for optical response calculations.\cite{Onida02,Cunningham18,Cunningham23}
Specifically, we here use an extension of the QS$GW$ method in which the screened Coulomb interaction is calculated including ladder diagrams, which has been dubbed
QS$G\hat{W}$,\cite{Cunningham23} and involves solving a BSE equation not just in the ${\bf q}\rightarrow0$ limit for optical response but for the mesh of ${\bf q}$ for which $W({\bf q},\omega)$ is needed. 
The muffin-tin-orbital basis functions  have atom centered 
spherical harmonic times smoothed  Hankel functions as envelope functions \cite{Bott98,questaalpaper}, which
are then replaced in its own and every other sphere by an expansion in spherical harmonics times radial solutions
of the Schr\"odinger equation  at a linearization energy and
its energy derivative which match in value and slope to the envelope functions.
This process is called augmentation. The expansion of the basis function centered  on one site about another is carried out by means of structure
constants. The basis set angular momentum cut-offs for first and second set of smoothed Hankel function basis sets were $\ell_{max}$=3,3 for Ga and O and  3,2 for Li.
Augmentations inside the spheres were carried out up to a higher cut-off
of  $\ell_{max}\le4$.
In the QS$G\hat{W}$ calculations 32 valence bands and 4 conduction bands were included in the active space in which the ladder diagrams are evaluated.
The PBEsol functional was used for the initial DFT calculations \cite{pbesol}. However, we did not minimize the structure within
this functional but used the experimental structural parameters and used the PBEsol only
as starting band structure  for the subsequent $GW$ calculations. 
In the QS$GW$ method, the energy dependent self-energy matrix $\Sigma_{ij}({\bf k},\omega)$
in the basis of the Kohn-Sham eigenstates $\psi_i$ is replaced by a Hermitian
$\tilde\Sigma_{ij}({\bf k})=\scriptstyle\frac{1}{2}\textstyle\Re{[\Sigma_{ij}({\bf k},\epsilon_i)+\Sigma_{ij}({\bf k},\epsilon_j)]}$ and this acts as a non-local exchange-correlation potential
whose difference  from the DFT exchange-correlation potential is added to the Kohn-Sham
Hamiltonian from which the $\Sigma$ is calculated in the next iteration until
convergence is reached. The self-energy is calculated explicitly up to
some maximum energy  $\omega_{max}=3.5$ Ry. The self-energy
matrix $\Sigma_{ij}({\bf k},\omega)$
is approximated by its diagonal above 3.0 Ry with an average value evaluated over the range $3.0<E<3.5$ Ry. This procedure is used to
facilitate interpolation of the $GW$ bands to other {\bf k}-points than
the mesh on which the self-energy is calculated explicitly. This is similar to a Wannier function
interpolation procedure where the muffin-tin orbitals themselves serve as  Wannier functions.
For a full description of these technical aspects of the method we refer the reader to Ref. 
\cite{Kotani07}.

\begin{figure}
  \includegraphics[width=8cm]{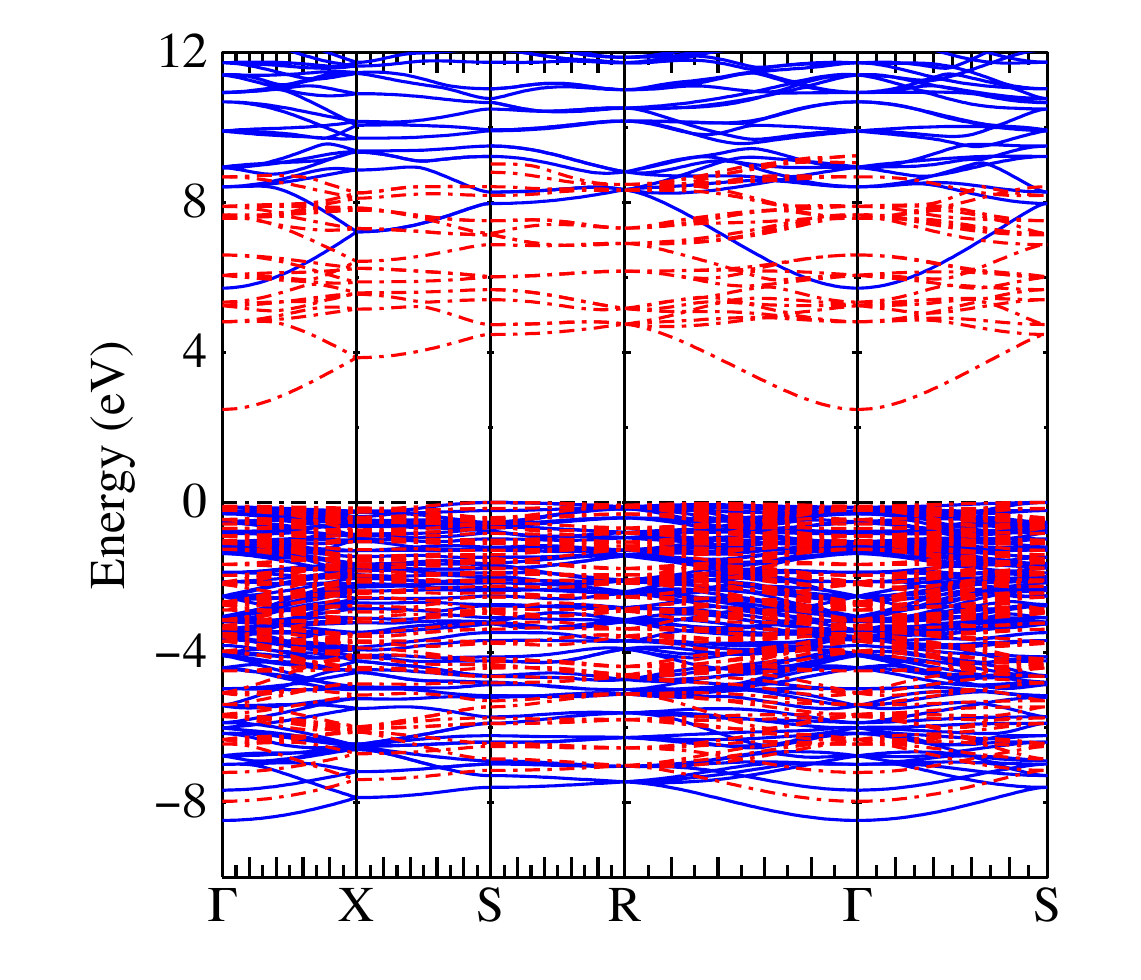}
  \caption{Band structure of LiGa$_5$O$_8$ in GGA (red dashed) and QS$G\hat{W}$ (blue solid)
    method.\label{figbnds} The Brillouin zone high-symmetry points follow the convention 
    for a simple cubic structure, $\Gamma=(0,0,0)$, $X=(\pi/a,0,0)$,
    $S=(\pi/a,\pi,a,0)$, (labeled $M$ in  \cite{Bilbao,Aroyo14}),  $R=(\pi/a,\pi/a,\pi/a)$}
\end{figure}
\begin{figure}
  \includegraphics[width=8cm]{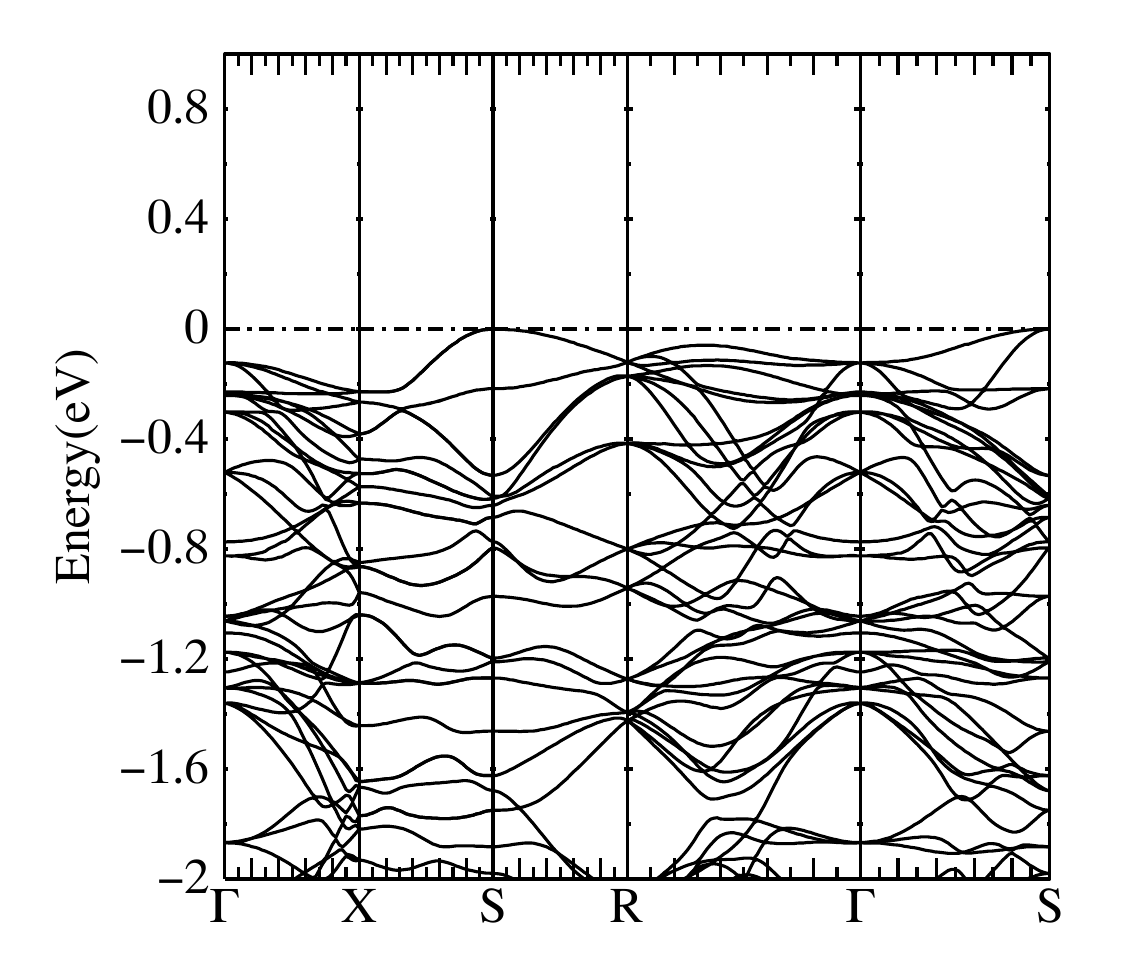}
  \caption{Zoom in near the valence band maximum of LiGa$_5$O$_8$ in QS$G\hat{W}$ method.
    \label{figzoom}}
\end{figure}

\section{Results}
The band structure of LiGa$_5$O$_8$  is shown in Fig. \ref{figbnds} as obtained
in the GGA and QS$G\hat{W}$ methods. The conduction band minimum (CBM) occurs at $\Gamma$ and the top valence band is very flat but with valence band maximum (VBM)
at $S$ as shown in Fig. \ref{figzoom}.  The gap is thus slightly indirect with indirect gap of 5.72 eV and direct gap at $\Gamma$ at 5.84 eV.
The gap is severely underestimated by GGA. Further details on the gap are given in Table \ref{tabgaps}.
The difference between the QS$G\hat{W}$ and GGA gap is 3.24 eV  and results from a downward shift of the VBM by 1.43 eV and 1.81 upward shift of the CBM.
These individual band edge shifts are obtained by directly comparing the DFT and the  QS$GW$ band edges relative to the common reference potential,
which in the LMTO method is set by an average of the potentials at the muffin-tin radii. It assumes that the self-consistent charge density and the potential apart from the $GW$-self-energy terms are the same in both cases, which is an excellent approximation.

From Table \ref{tabgaps} we can see that the differences between the QS$GW$ and QS$G\hat{W}$ gaps in this material is very small. This is somewhat unusual. In most materials studied so far,\cite{Cunningham23}  including the ladder diagrams  reduces the
$GW$ self-energy by about 10-20 \%. For LiGaO$_2$, we  already found a rather small reduction (5 \%) of the  self-energy when including ladder diagrams but 
here the reduction is only 1 \%. However, the small effect of ladder diagrams  found here may be a result of including an insufficient number of bands in the active space, \ie the space spanned by the
$N_v\times N_c\times N_k$ vertical excitation basis states in the two-particle Hamiltonian solved in the BSE step.
$N_v$ is the number of occupied states included, $N_c$ the number of empty conduction bands included and $N_k$ the number of {\bf k}-points. 
We estimate the reduction of the self-energy by
\begin{equation}
  1-\frac{E_g^{QSG\hat W}-E_g^{DFT}}{E_g^{QSGW}-E_g^{DFT}}.
\end{equation}
From the study of standard semiconductors, the guideline is to include the upper set of anion-$p$-like valence bands and
a number of conduction bands up to an equivalent energy above the CBM.  
Unfortunately,  for LiGa$_5$O$_8$ including all O-$2p$ bands would require 96 valence bands and including at least one band per Ga in the conduction band would require 20 conduction bands. This presently exceeds the memory allocation  requirements we can afford.
To check the importance of this convergence issue, we revisited 
our previous calculations of LiGaO$_2$ but with a higher number of
conduction bands. Instead of using all 52 occupied bands (20 Ga-$d$ like, 8 O-$2s$ and 24 O-$2p$)
and only 4 conduction bands which was the default used in \cite{Dadkhah23} we now use 24 valence
bands and 12 conduction bands. This gives a gap of 6.737 eV instead of 7.016 eV in \cite{Dadkhah23}. With a QS$GW$ gap of 7.218 eV and GGA gap of 3.314 eV, the $\Sigma_{QSG\hat{W}}$ is then estimated to be reduced relative to $\Sigma_{QSGW}$
by 12 \% instead of only 5 \%. 
It is important to note that while this reduces the quasiparticle gap of LiGaO$_2$ by about 0.3 eV, the same reduction of the screened Coulomb interaction also affects the exciton binding energy. Calculating the exciton gap accurately requires  a denser {\bf k} mesh. However,  for these low lying excitations we can fortunately use a smaller $N_v$ and $N_c$ but need a larger $N_k$ and
an extrapolation as function of $N_k$. This is different from the calculation of the ladder diagrams which involves the overall
effect of the screening and requires  a higher $N_c$. Using a similar extrapolation procedure as in \cite{Dadkhah23} for the excitons
we then find an extrapolated exciton gap of 6.4 eV. This is   in excellent agreement with  the result obtained in \cite{Dadkhah23}. However,
to compare with experiment, we also need to include a zero-point motion electron-phonon coupling correction, which was
there estimated to be $-0.4$ eV. This finally gave excellent agreement with the experimental value of 6.0 eV for the  lowest exciton gap.\cite{Tumenas17} In conclusion, the main result of that paper of an exciton gap at 6.0 eV and the qualitative analysis of the exciton
states presented in that paper holds when we here use a better converged QS$G\hat{W}$ calculations.
However, the quasiparticle gap and the exciton binding energy are both  reduced by about 0.3 eV. So, it does represent a correction to those previous results but one which is not
so easy to test experimentally as no independent measurements of the exciton binding energy or the quasiparticle gap
are available but  only the resulting optical gap or exciton energy. 
 This re-analysis of the LiGaO$_2$ case suggests that the QS$G\hat{W}$ gap in Table \ref{tabgaps} is still overestimated.
Using a reduction factor due to ladders of about 12 \% as suggested by the re-evaluation of LiGaO$_2$ we would obtain a indirect gap of 5.4$\pm0.1$ eV and direct gap at $\Gamma$ of 5.5 eV.

\begin{table}
  \caption{Band gaps of LiGa$_5$O$_8$  in eV}
  \begin{ruledtabular}
  \begin{tabular}{lccc}
    & GGA & QS$GW$ & QS$G\hat{W}$ \\ \hline
    indirect ($S-\Gamma$) & 2.48 &5.75 & 5.72 \\
    direct $\Gamma$ &  2.58 & 5.87 & 5.84 \\
  \end{tabular}
  \end{ruledtabular} \label{tabgaps}
\end{table}

\begin{figure}[h]
  \includegraphics[width=8cm]{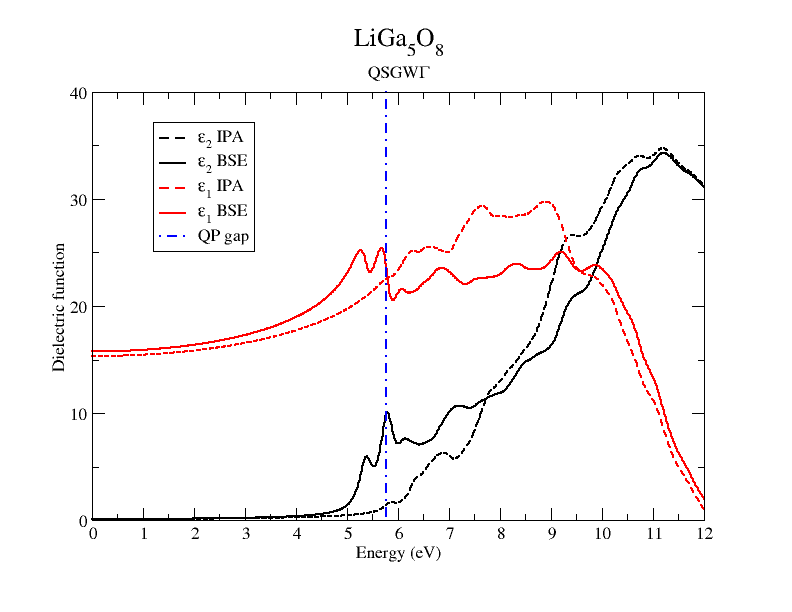}
  \caption{Real and imaginary part of the dielectric function.\label{figeps}}
\end{figure}

\begin{figure}[h]
  \includegraphics[width=8cm]{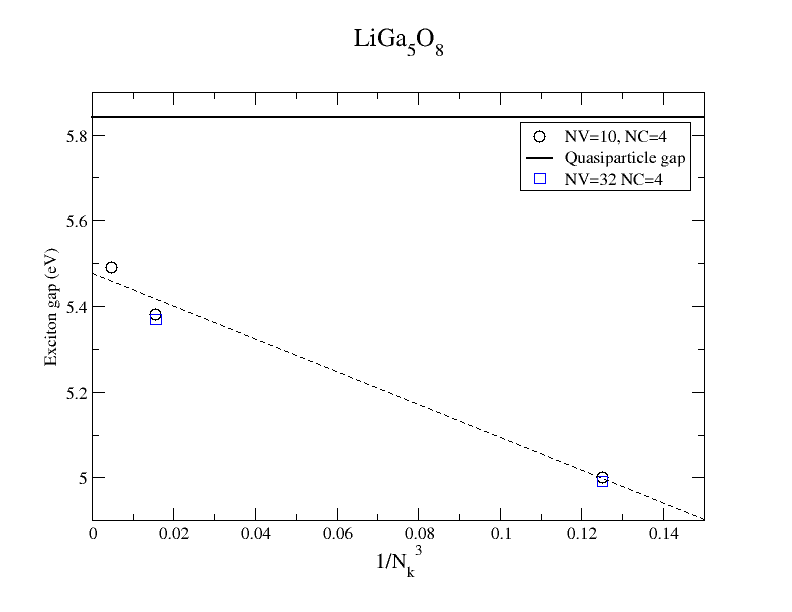}
  \caption{Exciton gap as function of {\bf k}-mesh. \label{figexgap}}
\end{figure}

The dielectric function is shown in Fig. \ref{figeps} in the independent particle approximation (IPA) and the
BSE approximation.  We can see a significant effect from including the electron-hole interaction effects and excitons with a large exciton
binding energy. We caution that  the dielectric constant $\varepsilon(\omega=0)$ obtained here is likely overestimated as is the intensity of $\varepsilon_2(\omega)$
because of difficulties in calculating the contribution of the self-energy to the velocity matrix elements but the peak positions should be accurate.  In this calculation we only used $N_k=2$
in a $N_k\times N_k\times N_k$ mesh for the BSE equation. To obtain the excitons more accurately, we calculated the lowest exciton for finer {\bf k}-meshes with $N_k=4,6$ but with  smaller  number of bands 10 valence bands instead of 32 while keeping 4 conduction bands.  We can see that reducing the number of valence bands   makes a difference of only 0.01 eV with slightly lower gaps obtained with more valence bands included.
Extrapolating the gap as function of $1/N_k^3$, \ie the total number of mesh points in the BZ, as we did in \cite{Dadkhah23} we find an extrapolated lowest exciton at 5.48 eV, close to our calculated gap for $N_k=6$.
This would amount to a exciton binding energy of $\sim$0.4 eV. 
The BSE calculations here only include direct transitions. Therefore we  deduce the
exciton binding energy with respect to the direct gap at $\Gamma$. One may expect a similar binding energy for an exciton related to the indirect gap which is $\sim$0.1 eV smaller.
Neither the quasiparticle gap nor the exciton gap here include electron-phonon coupling renormalization.
In LiGaO$_2$, the latter was found to be $-0.3\pm0.1$ eV. A similar value is expected here and
would  reduce our direct gap of Table \ref{tabgaps} from 5.87 to 5.6$\pm$0.1 eV (quasiparticle ) and 5.2.$\pm$0.1 eV (exciton) and another 0.1 eV lower for the indirect quasiparticle and exciton gaps.
Both the quasiparticle gap and exciton binding energy could be somewhat lower if we assume a larger
reduction of $\hat{W}$ relative to $W$ due to ladder diagrams but as mentioned earlier for LiGaO$_2$,
we expect these two errors to compensate each other and give a similar optical gap. 
This gap is slightly higher than the gap of 5.0 eV reported \cite{Mahitosh22} for $\gamma$-Ga$_2$O$_3$.

\begin{figure*}
 (a) \includegraphics[width=8cm]{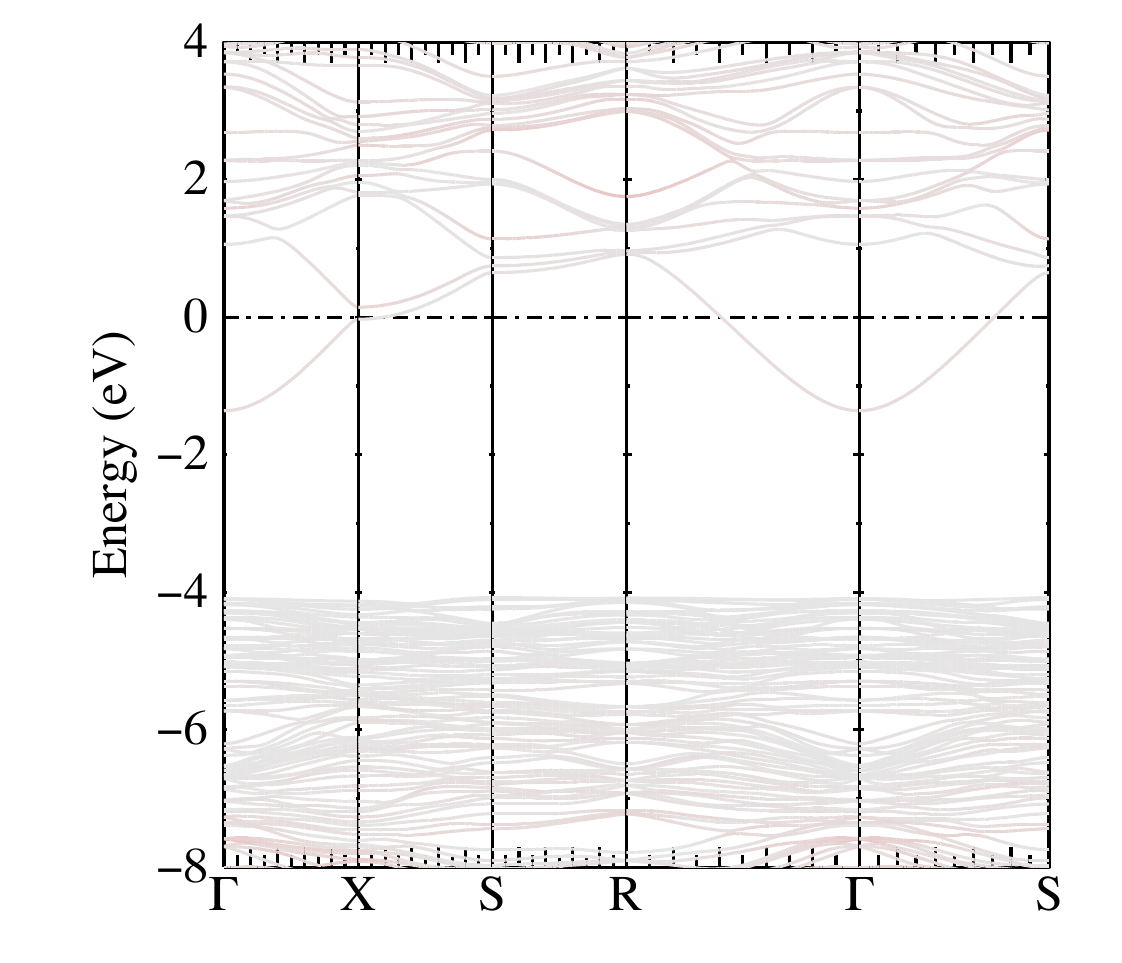}
  (b)\includegraphics[width=8cm]{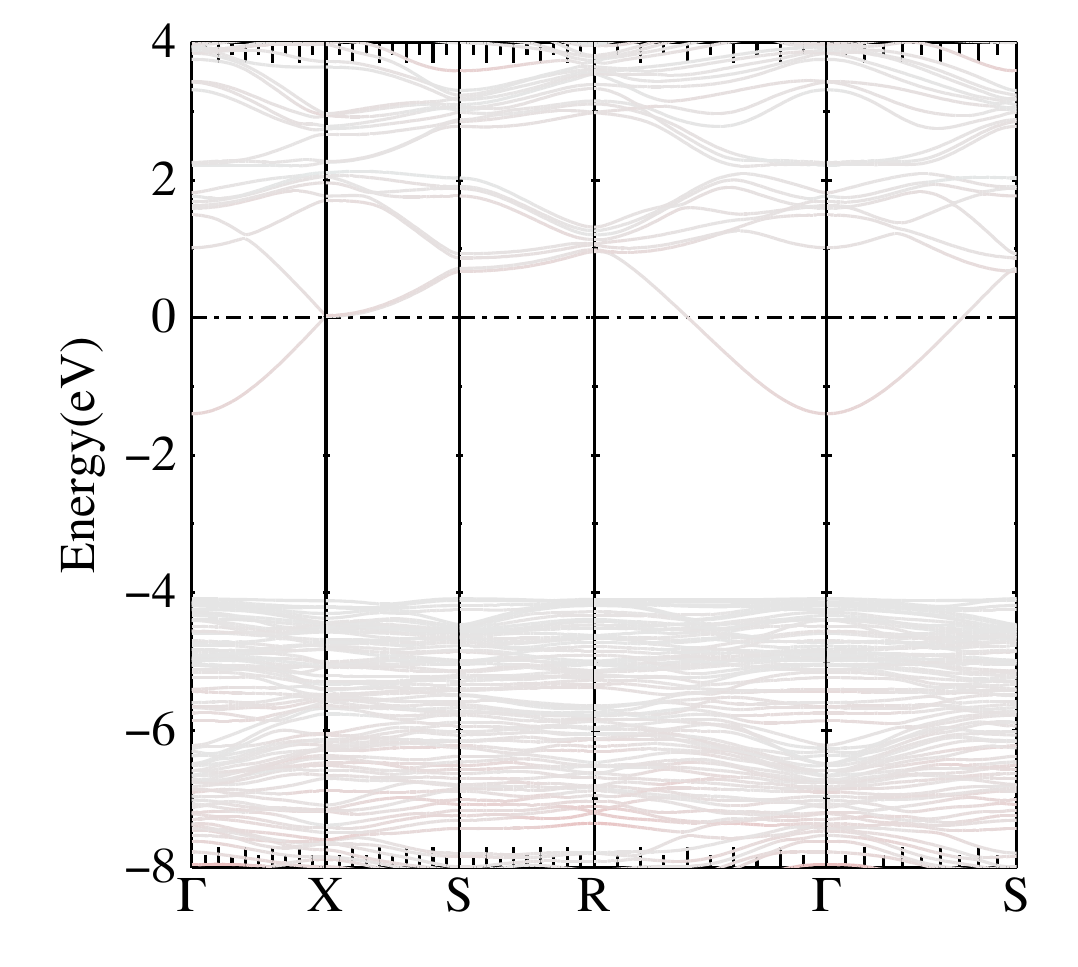}
  \caption{Band structure of SiLi$_4$Ga$_{19}$O$_{32}$ with Si in tetrahedral site (a), octahedral site (b). Red indicates the Si-dopant contribution to the bands, bands without Si are light grey.\label{figsit}}
\end{figure*}

\begin{figure}
  (a)\includegraphics[width=6cm]{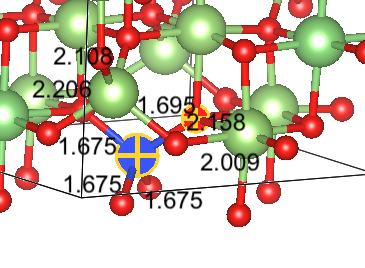}
  (b)\includegraphics[width=6cm]{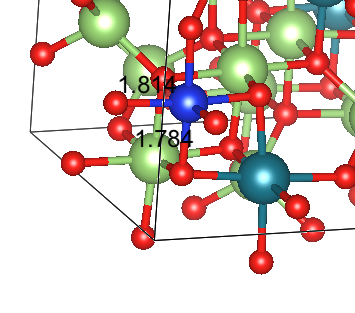}
  \caption{Bond lengths (\AA) near Si relaxed structure for: (a) tetrahedral, (b) octahedral site. The blue sphere is Si,  the green Ga and the teal one  is Li.
    The longest Si-O bond is indicated by the crosses in (a).\label{figbonds}}
\end{figure}
To evaluate the $n$-type dopability of the system, we replaced one Ga atom
in the 56 atom cell  with Si. Fig.\ref{figsit}(a)
shows the band structure for the tetrahedral Si site after relaxation of the atomic positions keeping the volume of the cell fixed
and in the GGA. One can see that
the extra valence electron essentially starts filling the conduction band
but no levels occur in the gap. The gap in fact slightly increased
from 2.58 eV in pure LiGa$_5$O$_8$ in the GGA to 2.74 eV. This is partly
due to the relaxation. Without relaxation the gap would have decreased slightly
to 2.39 eV. The Si contribution to the bands in this figure is indicated by the red color while the background color of the bands without Si contribution is grey.
The relaxation is significant. As can be seen in Fig. \ref{figbonds}, the nearest neighbor O move inward toward
the Si but with also a distortion  and symmetry lowering making one Si-O bond longer than the other three. The bond lengths change from 1.776 \AA\ for the Ga-O tetrahedral bond
to 1.675 \AA\ (3 bonds) and 1.695 \AA\ (one Si-O) bond.  The  Ga atoms bonded to these nearest neighbor O atoms then have a larger bond length again by about 0.1 \AA\ 
than the original  octahedral Ga-O bond length of 2.05 \AA.
  In spite of these relaxations there clearly is no deep level formation.
This indicates that Si would be a shallow donor and could lead to $n$-type doping.

We have also calculated Si on an octahedral site with similar results. First, we find that
the relaxed total energies were within error bar indistinguishable. So Si is likely to occupy either octahedral or tetrahedral sites with equal likelihood. Secondly, we  find no
states in the gap in either case, indicating that Si is a shallow donor on both sites as seen in
Fig. \ref{figsit}(b). Similar to the tetrahedral case, we find a slight increase in gap due to the Si replacement to 2.688 eV in GGA.  Two of the octahedral bonds around Si are 1.814 \AA, two are 1.864 \AA, and two are 1.784\AA.
The Ga-O octahedral bonds in the host are 1.982 \AA. Thus, there is again an inward relaxation of nearby O toward the Si.

While we have here only studied the Si doping via the GGA band structure, we
expect that the shallow donor character will stay valid at the
QS$GW$ level. A similar situation was found  recently for LiAlO$_2$.\cite{Popp22}

Of course, a full evaluation of doping possibilities will also require a study of native defects and compensation issues. 
Work along those lines is in progress and will be published elsewhere.

\section{Conclusion} In summary, 
LiGa$_5$O$_8$ in the spinel type cubic structure is a good candidate UWBG semiconductor. It is predicted to have slightly higher gap than $\gamma$-Ga$_2$O$_3$
of about 5.3 eV but not as high as LiGaO$_2$ and appears to be $n$-type dopable
by Si on either its tetrahedral or octahedral Ga site.  Its cubic structure
may have advantages.

\acknowledgements{This work made use of the High Performance Computing Resource in the Core Facility for Advanced Research Computing at Case Western Reserve University and was supported by the U.S. Air Force Office
  of Scientific Research (AFOSR) under grant no. FA9550-22-1-0201. I thank Hongping Zhao for communicating her results on epitaxial growth of LiGa$_5$O$_8$ prior to publication, which initiated my interest in this material.}

{\bf Data Availability}:
The data that supports the findings of this study are available within the article.

{\bf Conflicts of Interest}
The author has no conflicts to disclose.

\bibliography{dft,gw,lmto,ligao2,liga5o8,bse,ga2o3}

\end{document}